\documentclass[aps,prb,twocolumn,groupedaddress,showpacs]{revtex4}

\usepackage{graphicx}
\usepackage{amsmath}

\begin{document}

\title{Pressure dependence of the oxygen isotope effect in YBa$_2$Cu$_4$O$_8$}

\author{N. Suresh$^1$, J.G. Storey$^2$, G.V.M. Williams$^1$ and J.L. Tallon$^{1,2}$}

\affiliation{$^1$MacDiarmid Institute, Industrial Research Ltd.,
P.O. Box 31310, Lower Hutt, New Zealand.}

\affiliation{$^2$School of Chemical and Physical Sciences, Victoria
University, P.O. Box 600, Wellington, New Zealand.}

\date{\today}

\begin{abstract}
We have carried out measurements of the pressure dependence to 1.2
GPa of the oxygen isotope effect on $T_c$ in the high-$T_c$
superconductor YBa$_2$Cu$_4$O$_8$ using a clamp cell in a SQUID
magnetometer. This compound lies close to, but just above, the
1/8$^{th}$ doping point where in La$_{2-x}$Sr$_x$CuO$_4$ marked
anomalies in isotope effects occur. Both isotopes show the same very
large pressure dependence of $T_c$ with the result that the isotope
exponent remains low ($\sim$0.08) but increases slightly with
increasing pressure. This is discussed in terms of stripe
suppression, a competing pseudogap and the effect of superconducting
fluctuations.
\end{abstract}


\maketitle

There is ongoing debate as to whether electron-phonon interactions
play an important role in the physics of high-$T_c$ superconductors
and especially in the pairing interaction between carriers. The
experimental situation remains ambiguous. An oxygen isotope effect
has been observed in the $E(k)$ dispersion of
Bi$_2$Sr$_2$CaCu$_2$O$_{8+\delta}$ using angle resolved
photoemission spectroscopy (ARPES)\cite{Lanzara}. Surprisingly, the
isotope effect was observed at deeper energies outside the range of
renormalization (``kink") effects near the Fermi level, $E_F$. These
are difficult experiments especially given the resolution limits at
that time. Recent higher-resolution studies\cite{LaserARPES} using
laser ARPES indicate a more conventional scenario with the isotope
effect only observed in the renormalization of the self energy near
$E_F$ and a shift in the kink energy of about 3 meV. It is clear
then that electron-phonon interactions are important in these
systems but whether they contribute to pairing is another question.

In support of a minimal role, the isotope effect in $T_c$ near
optimal doping is found to be small\cite{Franck}, and though it
grows sharply with underdoping this is explicable in terms of the
effect of a competing normal-state pseudogap reducing the order
parameter towards zero while the spectral gap $\Delta$ remains
finite\cite{Williams}. With progressive underdoping this gives an
increasing (diverging as $T_c\rightarrow 0$) isotope effect in $T_c$
while that in $\Delta$ remains relatively unchanged. On the other
hand Lee {\it et al.}\cite{Lee} have observed a full oxygen isotope
exponent in a peak in $d^2I/dV^2$ in scanning tunneling spectroscopy
(STS) measurements when the spectra are referenced to the locally
observed gap. In STS, local-gap referencing is important because the
gap is found to be spatially variable due to some
yet-to-be-established inhomogeneity. Lee {\it et al.} have therefore
identified a possible bosonic energy scale, $\Omega$, for
Eliashberg-like interactions with the electronic system that could
be associated with the pairing. The exponent
$-\partial\ln\Omega/\partial\ln M$ was found to be close to 0.5
indicating a pure phononic interaction.

There are further complications with high-$T_c$ superconductors
(HTS). Firstly, around a hole concentration of $p$=0.125, at the
so-called 1/8$^{th}$ point, there occurs a very marked anomaly in
the isotope effects for both $T_c$\cite{Crawford,Tallonrho} and the
superfluid density $\rho_s$\cite{Tallonrho} in
La$_{2-x}$Sr$_x$CuO$_4$. This reflects the presence of
stripes\cite{Tranquada} or perhaps a checkerboard
structure\cite{Davis} in which the spins and charges spatially
separate and order. This results in a strong coupling of the
electronic system to the lattice and a large resultant isotope
effect, well above the canonical behavior expected for a competing
pseudogap\cite{Tallonrho}.

Secondly, the underdoped cuprates exhibit a remarkably large
pressure dependence of $T_c$ that is yet to be fully
understood\cite{Bucher,TallonP,Schilling}. One effect of pressure is
to induce charge transfer, doping additional carriers into the
conduction band. Pressure thus offers a mechanism to traverse the
phase diagram\cite{Schilling}. In lightly underdoped cuprates
($0.125<p<0.16$) it is found that $T_c(P)$ rises with pressure to a
maximum then falls\cite{Schilling}. Surprisingly, the maximum can be
much greater than the value of $T_{c,max}$ found at ambient
pressure. For example at ambient pressure YBa$_2$Cu$_4$O$_8$ has a
large d$T_c$/d$P$ coefficient of $\approx5.5$
K/GPa\cite{Bucher,TallonP} and $T_c$ rises to a maximum of
108K\cite{Eenige} at somewhere between 9 and 12 GPa before falling
at higher pressure. Y$_2$Ba$_4$Cu$_7$O$_{15-\delta}$, which is less
underdoped, also has a large pressure coefficient\cite{TallonP} of
d$T_c$/d$P$ = 4.1 K/GPa. On the other hand, for optimal and
overdoped cuprates d$T_c$/d$P$ is substantially reduced and becomes
negative with overdoping.

These effects have been quantified by adopting the commonly used
parabolic phase curve\cite{Presland}
\begin{equation}
T_c(P) = T_{c,max}(P) [1-82.6(p(P)-0.16)^2] \label{Tc}
\end{equation}
where both $T_{c,max}$ and $p$ are pressure dependent, and
$dp/dP\approx0.0055$ holes/Cu/GPa\cite{Schilling}. Such an analysis
results in a value of $dT_{c,max}/dP$ which is strongly pressure
dependent, being very large in the lightly underdoped region
($\geq3$ K/GPa), and becoming rather small in the overdoped region.
However, the model clearly breaks down in the more underdoped region
($0.05<p<0.13$) where $T_c(P)$ rises slowly with pressure and to a
maximum which little exceeds the ambient value\cite{Schilling}. The
continued use of equ.~\ref{Tc} in this region would require a large
negative value of $dT_{c,max}/dP$ and an abrupt crossover of
$dT_{c,max}/dP$ to large, positive values occurring close to
$p$=1/8. Along with the anomalously large isotope exponent in the
superfluid density\cite{Tallonrho} this is perhaps the clearest
indication of a discontinuity in the phase diagram occurring near
1/8$^{th}$ doping. The uniaxial stress dependence of $T_c$ in
(Y$_{1-y}$Ca$_y$)Ba$_2$Cu$_3$O$_x$ single crystals using
high-resolution thermal expansion measurements also shows a marked
discontinuity at 1/8$^{th}$ doping\cite{Meingast}.

But this leads us to another problem. In the lightly underdoped
region $T_{c,max}$ has this unexpectedly large positive pressure
coefficient. The curious thing is that the use of bond valence sums
to characterize bond stresses leads to the conclusion that
$T_{c,max}$ has a negative pressure coefficient\cite{BVS,Ionsize}.
This is effectively confirmed across the series
RBa$_2$Cu$_3$O$_{7-\delta}$ where, as R is increased in size from Yb
to La, $T_{c,max}$ increases from 92K to 101K by the ion size
effect\cite{Ionsize}. Based on these considerations the magnitude of
$T_{c,max}$ should decrease with pressure - it should have a
negative pressure coefficient, while the opposite is observed. This
contradiction is yet to be resolved and it may need characterization
of the pressure dependence of the competing pseudogap to progress
our understanding of this issue - see below. In the absence of data
on the pressure dependence of the pseudogap we will consider a
further issue, the role of fluctuations and their response to
pressure.

\begin{figure}
\centerline{\includegraphics*[width=85mm]{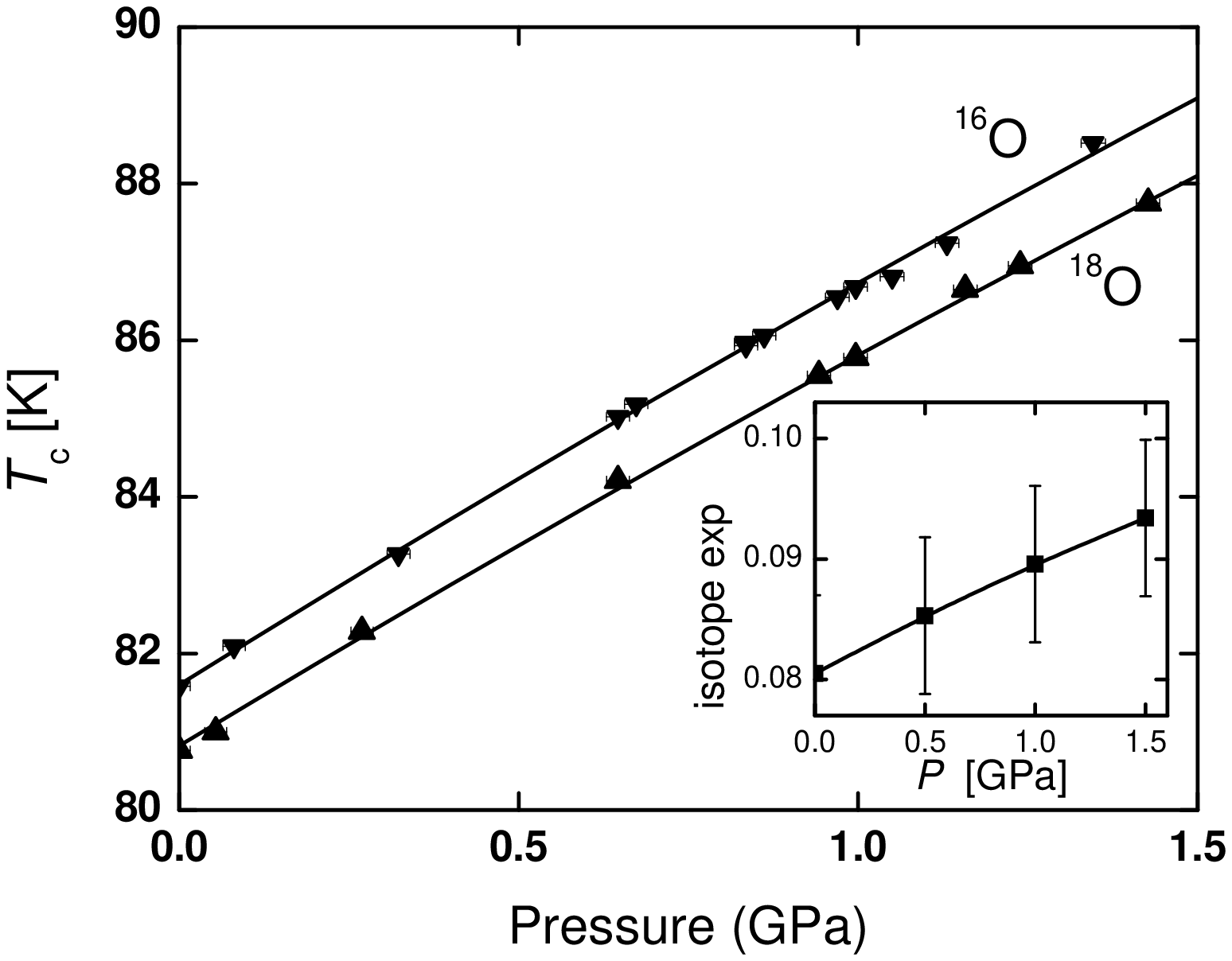}} \caption{\small
The pressure dependence of $T_c$ for YBa$_2$Cu$_4$O$_8$ with
$^{16}$O and $^{18}$O oxygen isotope exchange. Inset: pressure
dependence of the oxygen isotope exponent, $\alpha(T_c)= -
\frac{M\times\Delta T_c}{\Delta M\times T_c}$.}
\end{figure}

There are thus anomalous pressure effects and anomalous isotope
effects in underdoped cuprates. In the present work we bring these
issues together to examine the pressure dependence of the oxygen
isotope effect in YBa$_2$Cu$_4$O$_8$ (Y-124). We are motivated by
the facts that (i) Y-124 lies very close to 1/8$^{th}$ doping with
an estimated hole concentration of $p$=0.13. It is therefore likely
to be influenced by stripe or checkerboard fluctuations if they are
present in this system; (ii) Y-124 lies in the underdoped region
where the pressure dependence of $T_c$ is unusually high; (iii) it
is sufficiently underdoped that the pseudogap plays a very important
role in governing its thermodynamic and transport properties; (iv)
it is rigidly oxygen stoichiometric and therefore oxygen isotope
exchange, ensuring consistency of oxygen content and doping state,
is straightforward; and (v) this material is one of the most
defect-free HTS materials known. In the event, we find nothing too
unexpected in the pressure dependence of the isotope effect. Both
isotopic forms of Y-124 exhibit much the same large pressure
dependence over the pressure range considered. The result is that
the isotope exponent remains small ($\approx0.08$) rising only very
slowly (and perhaps not significantly) with $P$.

Samples were prepared by standard solid-state reaction at 945
$^\circ$C and 60 bar oxygen pressure as described
previously\cite{Williams}. A 12 mm diameter pellet was cut in half
and one half was annealed in 98\% $^{18}$O and the other in pure
$^{16}$O using gold baskets placed in adjacent narrow fused quartz
tubes. The samples were annealed at 760$^\circ$C for six hours at
0.95 bar pressure. The anneal was repeated four times with a new
charge of O$_2$ gas each time. This resulted in a mass change in the
$^{18}$O sample consistent with 95\% oxygen exchange. The nearly
complete isotope exchange was confirmed by the shift in oxygen
vibrational modes determined from Raman spectra\cite{Williams}.

For each sample the superconducting transition temperature, $T_c$,
was determined as a function of pressure in a Quantum Design MPMS
SQUID magnetometer using a miniature home-built piston clamp cell.
The cell, with dimensions of 8.8mm diameter and 65mm length, was
made from non-magnetic Be-Cu(Mico Metal 97.75\% Cu, 2\% Be) with
cobalt-free tungsten-carbide pistons (from Boride). The pistons are
lightly tapered using electric-discharge machining\cite{Walker}. The
sample was loaded in a 2.67 mm diameter 9 mm long Teflon capsule
along with Fluorinert FC70 and FC77 mixed in 1:1 ratio as a
cryogenic hydrostatic pressure medium. A small piece of high purity
Pb wire was included as a pressure calibrant. To apply pressure the
cell was preloaded before clamping at room temperature using a
laboratory press with calibrated digital pressure gauge (Ashcroft
Model 2089, 0.05\% accuracy). The pressure in the sample was
measured from the reported shift in $T_c$ of Pb at zero
field\cite{Eiling,Suresh}. In the course of this work we found that
the pressure dependence of $H_c$ had not been reported for Pb so we
carried out an extensive investigation of $H_c(P,T)$ from which we
determined the basic thermodynamic functions: the electronic
entropy, specific heat coefficient, compressibility and thermal
expansion coefficient\cite{Suresh}.

Zero-field cooled temperature sweeps were made at 2.5 mT to measure
the diamagnetic magnetization and hence the onset of
superconductivity. $T_c$ values are plotted in Fig. 1 for the two
isotope exchanged Y-124 systems. The increase in $T_c$ with pressure
reveals a small but clear quadratic curvature consistent with
measurements over a much broader pressure range\cite{Griessen}.
Quadratic fits gave the following results with $^{18}T_c(P) = 80.82
+ 5.25P - 0.26P^2$ for the $^{18}$O sample and $^{16}T_c(P) = 81.6 +
5.4P - 0.27P^2$ for the $^{16}$O sample. The value of d$T_c$/d$P$ =
5.4 K/GPa in the latter case is similar to previous
reports\cite{Bucher,TallonP,Eenige} while the implicit value at
$P=0$ of $\alpha \equiv$ -d$\ln T_c$/d$\ln M$ = 0.0805 is also
comparable to previous reports at ambient pressure\cite{Williams}.
Moreover, extrapolation of the quadratic fit results in a maximum
$T_c$ value of 108K occurring at 10 GPa. This is very consistent
with the values reported from a collection of diamond anvil studies
at higher pressures\cite{Griessen}.

The inset to Fig. 1 shows the values of $\alpha$ deduced from the
quadratic fits at four different pressures and corrected for the
95\% isotope exchange. This shows only a small increase with
pressure (which might, however, be statistically insignificant).

How then are we to understand the anomalously high pressure
coefficient in $T_c$ together with the small isotope exponent that
grows only slightly with increasing pressure?

Firstly, let us consider what might be expected. Pressure generally
transfers extra holes onto the CuO$_2$ planes\cite{Schilling}.
Because the isotope effect is usually observed to decrease with
increasing doping one might thus expect the isotope effect to
decrease with pressure. Moreover, an increase in doping would tend
to move the sample further away from 1/8$^{th}$ doping and hence
away from any anomaly associated with stripe
instability\cite{Crawford}. This also would tend to decrease the
isotope effect\cite{Tallonrho}. For Y-124, however, these effects
will be small because it already sits close to the plateau in
$\alpha(p)$\cite{Pringle}. More important is the large pressure
coefficient in $T_c$ and this does not seem explicable in terms of a
simple doping effect, requiring as it does an anomalously large
value of $dT_{c,max}/dP$ (and of opposite sign to what is
expected)\cite{BVS,Ionsize}.

One has also to consider the possibility that the tendency to stripe
formation is enhanced under pressure. La$_{2-x}$Sr$_x$CuO$_4$ is
generally recognized to be the most inclined to stripe formation of
the common high-$T_c$ superconductors. Near $p=1/8$ this material
exhibits a very large isotope effect in both $T_c$ and in the
superfluid density, $\rho_s$, departing severely from the canonical
behavior expected for a pseudogap competing with
superconductivity\cite{Tallonrho}. This indicates the strong
coupling of the electronic system to the lattice in this region
where stripe or checkerboard inhomogeneity occurs. Interestingly the
Y-123 system remains on the canonical line of $\alpha(\rho_s)$
versus $\alpha(T_c)$ showing no apparent stripe-derived
anomaly\cite{Tallonrho}. But we should not conclude from this that
``stripe" correlations are irrelevant in Y-123 and Y-124. The
$T_c(p)$ phase curve for (Y,Ca)-123 is almost identical to that for
La$_{2-x}$Sr$_x$CuO$_4$ with the ``60K plateau" in Y-123 coinciding
with the $T_c$ anomaly in La$_{2-x}$Sr$_x$CuO$_4$ at 1/8$^{th}$
doping\cite{Tallonstripe}. This feature in Y-123 is often thought to
be associated with oxygen (``ortho-II") ordering but substitution of
La for Ba, and Ca for Y shows that it is pinned to $p=1/8$,
independent of oxygen content\cite{Tallonstripe}. Interestingly,
Y-124 (doped with La) also exhibits a 60K
plateau\cite{Tallonribbons}. These features are almost certainly
associated with short-range stripe correlations.

The CuO$_2$ planes in La$_{2-x}$Sr$_x$CuO$_4$ experience an in-plane
compression in comparison with Y-123 and other higher $T_c$
systems\cite{BVS} resulting in a higher antiferromagnetic exchange
interaction, $J$. If the shorter Cu-O bondlength and enhanced value
of $J$ is conducive to stripe formation near $p=0.125$ then this
would explain the strong stripe tendency in La$_{2-x}$Sr$_x$CuO$_4$.
Following this line of reasoning, the effect of pressure on Y-124
might be to enhance the tendency to stripe formation. Because it
resides close to $p=0.125$ the increasing isotope exponent could
signal an increasing tendency to stripe formation, even in this
compound. But stripe development reduces $T_c$ and, in this case,
one would expect a negative pressure coefficient of $T_c$. In fact
the opposite is found to be the case. Pressure suppresses stripes in
La$_{2-x}$Ba$_x$CuO$_4$, narrowing the domain around $p$=1/8 in
which $T_c$ is diminished\cite{Ido} and therefore resulting in a
large positive value of $dT_c/dP$.

We are thus presented with a double anomaly. Correlation of
compression effects across the cuprates, revealed for example using
bond valence sums, would suggest that (i) $T_{c,max}$ should be
diminished by pressure, and (ii) stripes should be enhanced by
pressure (which in turn would also tend to decrease $T_c$). Neither
of these is true. We probably, therefore, need to look to other
possible effects to explain these results. What, for example, can
explain the very large pressure coefficient in $T_c$ (for either
isotope), leading to  $T_{c,max}$ as high as 108K in Y-124. We
consider two further possibilities: the role of (i) the pseudogap
and (ii) fluctuations.

In the underdoped and optimally-doped regions the pseudogap competes
with superconductivity, reducing $T_c$, $\rho_s$, the jump in
specific heat at $T_c$ and the condensation energy, $U_0$. This
causes an abrupt crossover from overdoped strong superconductivity
to underdoped weak superconductivity at $p = p_{crit} = 0.19$
holes/Cu. A $P$-dependent rise in $T_{c,max}$ could thus be
associated with a $P$-dependent decrease in the pseudogap (or
alternatively a $P$-dependent shift of $p_{crit}$ to lower doping).
We consider this to be unlikely in view of the fact that the
pseudogap has been associated with short-range AF spin
fluctuations\cite{TallonLoram,TallonSG}. An increase in pressure
should increase the pseudogap energy scale, $E_g$, hand in hand with
the pressure-induced increase in $J$\cite{Aronson,Maksimov}.
Similarly, one would expect pressure to shift $p_{crit}$ to higher
doping. These are ideas which we propose to test using
pressure-dependent thermopower and Raman spectroscopy studies.
However, as we have said - we expect that pressure effects on the
pseudogap probably cannot account for the large positive value of
$dT_c/dP$ in lightly underdoped cuprates.

This leaves us with fluctuations. In fact, we have formed the view
that, in optimal and underdoped cuprates, Gaussian fluctuations are
effective in significantly reducing $T_c$ below the mean-field
value. Evidence for this has been derived from a detailed
fluctuation analysis of the specific heat anomalies in
YBa$_2$Cu$_3$O$_{7-\delta}$ and
Bi$_2$Sr$_2$CaCu$_2$O$_{8+\delta}$\cite{Tallonfluc}. This leads to a
monotonically decreasing mean-field transition temperature,
$T_c^{mf}$, as a function of doping, along with the usual parabolic
$T_c(p)$ - the downturn at low doping being due to strong
fluctuation effects. If so, then the effect of pressure may simply
be to increase the interlayer coupling thus reducing the effect of
fluctuations and raising $T_c$ closer to the mean-field $T_c$ value.
The pressure derivative would increase sharply with decreasing
doping, as observed. We intend to probe such effects by
investigating the pressure-dependence of the fluctuation
para-conductivity. The isotope effect would take the value of the
isotope effect in the mean-field $T_c^{mf}$ value which in turn
would take its value from the isotope effect in $\Delta$, the
superconducting energy gap. This could naturally explain our
observation of a very strong pressure dependence of $T_c$ and a weak
pressure dependence of the isotope effect.

In conclusion, for YBa$_2$Cu$_4$O$_8$ we find a very strong pressure
dependence of $T_c$ for both $^{16}$O and $^{18}$O isotopes in
combination with a zero, or small, pressure-induced increase in the
oxygen isotope effect. These two effects are difficult to reconcile
with simple pressure-induced charge transfer and bond compression
which increases the magnitude of $J$. We surmise that the effect of
pressure is to reduce the Gaussian fluctuations which depress $T_c$
below its mean-field value. Some tests of this hypothesis are
proposed.

\end{document}